\documentclass[pss,fleqn]{w-art}
\usepackage{times}
\usepackage{amsmath}
\usepackage{w-thm}
\usepackage{bm}
\usepackage{graphicx}
\usepackage{amssymb}

\begin{document}


\title{Diffusion thermopower of a $p-$type Si/Si$_{1-x}$Ge$_x$
 heterostructure at zero magnetic field}
\author{Tran Doan Huan}
\address{Institute of Engineering Physics, Hanoi
University of Technology, 1 Dai Co Viet Rd., Hanoi, Vietnam}
\author{Nguyen Phuc Hai\footnote{Corresponding author,
email: hainp-iep@mail.hut.edu.vn}}
\address{Faculty of Engineering, Katholieke University of Leuven,
Kasteelpark Arenberg 44, B-3001 Heverlee (Leuven), Belgium}

\begin{abstract}
We calculate the diffusion thermopower of the degenerate
two-dimensional hole gas in a $p-$type Si/Si$_{1-x}$Ge$_x$ lattice
mismatched heterostructure at low temperatures and zero magnetic
field. The effects of possible scatterings, e.g. remote impurity,
alloy disorder, interface roughness, deformation potential, and
random piezoelectric on the hole mobility and the diffusion
thermopower are examined. Calculated results are well fitted to the
experimental data recently reported. In addition, we predict a
possibility for the diffusion thermopower to change its sign as the
SiGe layer thickness changes, the effect has not been discussed yet.

\end{abstract}

\maketitle

\section{Introduction}
In a recent experiment, the thermopower of the two-dimensional (2D)
hole gas in a $p-$type Si/Si$_{0.88}$Ge$_{0.12}$ heterostructure
(HS) were reported \cite{pos04}. At very low temperature, the total
thermopower $S$, approximated by the diffusion thermopower $S^{\rm
d}$, was explained by a \textit{phenomenological} model without
calculations starting from the microscopic level \cite{pos04}. Since
many theoretical studies \cite{kar90, kar91} of the diffusion
thermopower are available, it may be interesting if one can use them
to examine the reported data.

Among the scatterings determining diffusion thermopower, impurity
doping and interface roughness are normally studied
\cite{kar90,kar91}. However, proper considerations for diffusion
thermopower due to deformation potential and alloy disorder, to the
best of our knowledge, are unavailable. Further, some assumptions
used in the existing studies are actually weak. First, the potential
barrier at the HS's interface is always assumed to be infinite
\cite{kar90, kar91}, although it is small, thus changes the
transport properties of the HS \cite{kar91, qth04}. Next,
deformation potential scattering limiting the 2D hole mobility is
always based on the idea that the holes and the electrons undergo
the same deformation potential \cite{sa00}. This assumption was
indicated \cite{qth04} to be invalid, and the properly derived
deformation scattering was found \cite {qth04} to strongly limit the
2D hole mobility. Consequently, one may interest in studying the
diffusion thermopower determined by this scattering. Lastly, a study
of the diffusion thermopower caused by random piezoelectric field,
the recently introduced scattering \cite{qth04}, is open. This
scattering was found \cite{kub05} to cause a sign change of the
diffusion thermopower in an $n-$type GaAs quantum well as the well
thickness varies. Even though piezoelectric scattering in SiGe alloy
is weak \cite{qth04}, a study of the diffusion thermopower
associated with it may be useful.

In this paper, we calculate the diffusion thermopower of the 2D hole
gas in a $p-$type Si/Si$_{0.88}$Ge$_{0.12}$ HS with finite barrier.
The scatterings to be considered are remote impurity, alloy
disorder, interface roughness, piezoelectric, and the properly
derived deformation potential. Calculated diffusion thermopower is
given in comparison with experiment \cite{pos04}. We also discuss
the possibility for the diffusion thermopower to change its sign as
the SiGe layer thickness varies.

\section{Diffusion thermopower}

The total thermopower $S$ is defined by ${\bf{\it{\cal E}}}=S\nabla
T$ in the presence of a temperature gradient $\nabla T$ and an
electric field $\bf{\it{\cal E}}$. At the limit of weak coupling,
the total thermopower $S$ of a degenerate 2D gas is $S=S^{\rm
d}+S^{\rm g}$, where $S^{\rm d}$ and $S^{\rm g}$ are the diffusion
and phonon-drag components, respectively \cite{pos04, kar90, kar91}.
At low temperature $T$, $S^{\rm d}\propto T$ \cite{pos04,kar90,
kar91}, while $S^{\rm g}$ has a more complex behavior depending on
the carrier-phonon scattering. For deformation potential scattering,
$S^{\rm g}\propto T^6$ \cite{fle97}, while for piezoelectric
scattering, $S^{\rm g}\propto T^4$ \cite{fle02}. In the former case,
$S^{\rm g}$ dominates as $T\leq 1\rm{K}$, while in the latter,
$S^{\rm d}$ dominates below $0.5\rm{K}$. Thus, as $T<0.5{\rm K}$,
one can consider $S^{\rm d}$ as the total thermopower, which can be
measured \cite{pos04}. For elastic scatterings, the diffusion
thermopower $S^{\rm d}$ of a 2D hole gas at zero magnetic field is
given by \cite{kar90, kar91}:
\begin{equation}
\label{Eq:Sd1} S^{\rm d}=\frac{\pi^2k_{\rm
B}^2T}{3|e|}\left[\frac{d\ln\sigma(E)}{dE}\right]_{E=E_{\rm F}},
\end{equation}
where $|e|$ is the hole charge, $k_{\rm B}$ Boltzmann constant,
$E_{\rm F}=\hbar^2k_{\rm F}^2/2m^*$ the Fermi energy, $k_{\rm
F}=\sqrt{2\pi p_{\rm S}}$ Fermi wave number, and $m^*$ the hole
effective mass. In Eq. (\ref{Eq:Sd1}), the $\sigma(E)$ is the
conductivity given by $\sigma(E)=p_{\rm S}(E)e^2\tau(E)/m^*$ with
$\tau(E)$ is relaxation time and $p_{\rm S}(E)=Em^*/\pi\hbar^2$ the
2D hole density \cite{kar90, kar91}. It is always assumed
\cite{kar90, kar91} that $\tau(E)\propto E^p$, so the Eq.
(\ref{Eq:Sd1}) can be rewritten by Mott formula \cite{kar90, kar91}:
\begin{equation}
\label{Eq:p1} S^{\rm d}=\frac{\pi^2 k_{\rm B}^2T}{3|e|E_{\rm
F}}(p+1),
\end{equation}
with the scattering parameter $p$ defined by \cite{kar90, kar91}:
\begin{equation}
\label{Eq:p2} p=\frac{E_{\rm F}}{\tau(E_{\rm
F})}\left[\frac{d\tau(E)}{dE}\right]_{E=E_{\rm F}}.
\end{equation}
The hole gas in our HS, as will be seen later, is expected to
undergo simultaneously remote impurity, alloy disorder, interface
roughness, deformation potential, and piezoelectric scatterings. In
this case, the total relaxation time $\tau(E)$ is given by the
Matthiessen's rule \cite{an82}:
\begin{equation}
\label{Eq:tautot} \frac{1}{\tau(E)}=\frac{1}{\tau_{\rm RI}(E)}
+\frac{1}{\tau_{\rm AD}(E)}+\frac{1}{\tau_{\rm IFR}(E)}
+\frac{1}{\tau_{\rm DP}(E)}+\frac{1}{\tau_{\rm PE}(E)},
\end{equation}
where $\tau_{\rm RI}(E)$, $\tau_{\rm AD}(E)$, $\tau_{\rm IFR}(E)$,
$\tau_{\rm DP}(E)$, and $\tau_{\rm PE}(E)$ are the relaxation times
due to the scatterings listed above, respectively.

It could be seen from the Eqs. (\ref{Eq:p1}) and (\ref{Eq:tautot}),
that the total diffusion thermopower $S^{\rm d}$ of a 2D system,
which undergoes more than one scatterings, is specified \textit{not
only} by the scattering strengths, \textit{but also} the energy
dependence of the relaxation time $\tau(E)$ at $E=E_{\rm F}$. In
terms of the corresponding autocorrelation functions $\langle
|U(\textit{\textbf{q}})|^2 \rangle$, they are expressed by
\cite{qth04,kub05}:
\begin{equation}
\label{Eq:tau} \frac{1}{\tau(E)}=\frac{1}{(2\pi)^2\hbar
E}\int_0^{2q} dq'\int_0^{2\pi}d\theta \,\frac{q'^2}{\sqrt{4q^2-q'^2}}
\frac{\langle |U(\textit{\textbf{q'}})|^2 \rangle}{\epsilon^2(q')},
\end{equation}
where $\textit{\textbf{q'}} =(q',\theta')$, $E=\hbar^2q^2/2m^*$ is
the energy corresponding to wave vector
$\textit{\textbf{q}}=(q,\theta)$. The dielectric function
$\epsilon(q)$ in Eq. (\ref{Eq:tau}) is given at zero temperature
within the random phase approximation by \cite{an82}
\begin{equation}
\label{Eq:epsi} \epsilon(q)=1+\frac{q_{\rm TF}}{q}F_{\rm
S}\left(q\right)\,[1-G(q)]\;\;\;\; {\rm for} \; q\leq 2k_{\rm F},
\end{equation}
with $q_{\rm TF}=2m^*e^2/\epsilon_{\rm L}\hbar^2$ is the inverse 2D
Thomas-Fermi screening length, $\epsilon_{\rm L}$ the dielectric
constant of the HS, and the function $G(q)=q/2\sqrt{q^2+4k_{\rm
F}^2}$ in Eq. (\ref{Eq:epsi}) allows for the local field corrections
associated with the many-body interactions of the 2D hole gas
\cite{an82}. The form factor $F_{\rm S}(q)$ is defined by
\cite{qth04, oku89}:
\begin{equation}
\label{Eq:Fs}
F_{\rm S}\left(q\right)=\int_{-\infty}^{+\infty}\!dz\!\int_{-\infty}^{+\infty}
\!dz'|\zeta(z)|^2|\zeta(z')|^2e^{-q|z-z'|},
\end{equation}
where $\zeta(z)$ and $\zeta(z')$ are the wave functions representing
the two holes having interaction (\ref{Eq:Fs}). An explicit
expression for $F_{\rm S}(q)$ will be given in the Eq.
(\ref{Eq:FS}), thus $\langle |U(\textit{\textbf{q}})|^2\rangle$ is
all needed for specifying $\tau(E)$.

\section{Autocorrelation functions for the scattering mechanisms}

Scattering by a random field can be specified by its autocorrelation
function in the wave vector space $\langle
|U(\textit{\textbf{q}})|^2\rangle$ \cite{an82}. Here the angular
brackets stand for an ensemble average over the fluctuations of the
2D Fourier transform of the random scattering field
$U(\textit{\textbf{q}})$, given by \cite{qth04, an82}:
\begin{equation}
\label{Eq:UFTW} U(\textit{\textbf{q}})=\int_{-\infty}^{+\infty}
dz\,|\zeta(z)|^2U(\textit{\textbf{q}},z).
\end{equation}
In our HS, the 2D hole gas is confined by a triangular potential
\cite{pos04,col97} located along the growth direction chosen as the
$z$ axis with the Si/SiGe interface locates at $z=0$. It has been
shown \cite{an82b} that for the finite barrier potential $V_0$, the
lowest subband may be very well described by the modified
Fang-Howard wave function \cite{an82, an82b, oku89}:
\begin{eqnarray}
\label{Eq:zeta} \zeta(z)=\left\{
\begin{array}{lll}
A\kappa^{1/2}e^{\kappa z/2}&&{\rm for}\;z<0,
\\
Bk^{1/2}(kz+c)e^{-kz/2}&&{\rm for}\;z>0,
\end{array}
\right.
\end{eqnarray}
in which $k$ and $\kappa$ are half of the wave numbers in the well
and barrier, respectively. Further, $A,\;B,\;c,\;\kappa$, and $k$
are the variational parameters to be determined by minimizing the
ground state energy connecting to $V_0$ \cite{an82b}. Detailed
expressions to be minimized could be found elsewhere \cite{qth04,
an82, oku89, an82b, quang05}.

Using Eq. (\ref{Eq:zeta}), the form factor $F_{\rm S}(q)$ in
(\ref{Eq:Fs}) can be given in term of $t=q/k$ and $a=\kappa/k$ by
\cite{qth04,oku89}
\begin{eqnarray}
\label{Eq:FS}
F_{\rm S}(t)=\frac{A^4a}{t+a}+2A^2B^2a\frac{2+2c(t+1)+c^2(t+1)^2}
{(t+a)(t+1)^3}
+\frac{B^4}{2(t+1)^3}\Bigl[2(c^4+4c^3+8c^2+8c+4)
\nonumber \\
+\,t\,(4c^4+12c^3+18c^2+18c+9)
+\,t^2(2c^4+4c^3+6c^2+6c+3)\Bigr].
\end{eqnarray}

Our HS is doped by an impurity sheet of thickness $L_{\rm I}$
supplying a scattering source \cite{pos04, col97}. This impurity
sheet follows a spacer layer of thickness $L_{\rm S}$ grown on the
top of the SiGe layer of thickness $L$ \cite{col97}. Thus, the first
autocorrelation function we have to specify should be \cite{an82,
an82b, quang05}:
\begin{equation}
\label{Eq:ACFMI} \langle|U_{\rm
RI}(\textit{\textbf{q}})|^2\rangle=\left(\frac{2\pi e^2}{\epsilon_L
q}\right)^2\frac{q}{q+q_{\rm I}}\int_{L+L_{\rm S}}^{L+L_{\rm S}
+L_{\rm I}}dz_in_{\rm I}(z_i) F_{\rm I}^2\left(q,z_i\right)
\end{equation}
where $F_{\rm I}(q,z_i)$ is the remote impurity form factor of a 2D
impurity sheet located at $z_i$ defined by \cite{an82, an82b}:
\begin{equation}
\label{Eq:FMI1} F_{\rm I}\left(q,z_i\right)=\int_{-\infty}^\infty dz
|\zeta(z)|^2e^{-q|z-z_i|}
\end{equation}
while $n_{\rm I}(z)$ is the impurity density at $z_i$, which is
$n_{\rm I}$ as $L+L_{\rm S} < z_i < L + L_{\rm S} + L_{\rm I}$ and
zero otherwise.

Further, it was indicated \cite{sch90} that at high impurity doping
level, the impurity distribution is not completely random. Due to
the Coulomb interaction among the charged impurities during the
sample growth, the impurity diffusion tends to diminish the
probability of large fluctuations of the impurity density
\cite{efr90}. The so-called impurity correlation effect can be taken
\cite{quang05, quang99} by adding to the autocorrelation function a
screening-like factor $q/(q+q_{\rm I})$ in term of the inverse
statistical-screening radius $q_{\rm I}$ defined by \cite{quang05,
quang99}
\begin{equation}
q_{\rm I}=\frac{2\pi e^2n_{\rm I}L_{\rm I}}{\epsilon_Lk_{\rm B}T_0}.
\end{equation}
In the above expression, $T_0\sim 1000K$ is the freezing temperature
of the impurity system \cite{quang05, quang99}.

By plugging Eq. (\ref{Eq:zeta}) into Eq. (\ref{Eq:FMI1}), and then
putting results into Eq. (\ref{Eq:ACFMI}), we obtain $\langle|U_{\rm
RI}(\textit{\textbf{q}})|^2\rangle$ as:
\begin{eqnarray}
\label{Eq:U2}
\langle|U_{\rm RI}(\textit{\textbf{q}})|^2\rangle
\displaystyle{=\left(\frac{2\pi e^2}{\epsilon_L q}\right)^2
\frac{n_{\rm I}}{k}\frac{q}{q+q_{\rm I}}\left\{\frac{1}{2}
\left[Q_1^2\left(\frac{q}{k}, s\right)-Q_1^2\left(\frac{q}{k}, d
\right)\right]+4B^2\left[Q_1\left(\frac{q}{k}, s\right)Q_2
\left(\frac{q}{k}, s\right)\right.\right.}\nonumber\\
\displaystyle{\left.\left.-Q_1\left(\frac{q}{k}, d\right)Q_2
\left(\frac{q}{k}, d\right)\right]+B^4\left[Q_3\left(\frac{q}{k},
s\right)-Q_3\left(\frac{q}{k}, d\right)\right]\right\}}
\end{eqnarray}
with $s=k(L+L_{\rm S}), d=k(L+L_{\rm S}+L_{\rm I})$ and the
auxiliary functions defined by:
\begin{equation}
Q_1(t,v)=\frac{e^{-tv}}{\sqrt{t}}\left\{\frac{A^2 a}{t+a}+B^2\left[\frac{c^2}{1-t}
+\frac{2c}{(1-t)^2}+\frac{2}{(1-t)^3}\right]\right\},
\end{equation}
\begin{equation}
Q_2(t, v)=\frac{e^{-v}\sqrt{t^3}}{1+t}\left[\frac{(c+v)^2}{t^2-1}
+\frac{2c(t-3)}{(t^2-1)^2}+\frac{2v(t^3-3t^2-t+3)}{(t^2-1)^3}
+\frac{4(t^2-2t+3)}{(t^2-1)^3}\right],
\end{equation}
\begin{eqnarray}
Q_3(t,v)=e^{-2v}t^2\left[\frac{3t^8-20t^6+66t^4-84t^2+163}{(t^2-1)^6}
+\frac{2(c+v)(3t^6-17t^4+41t^2-91)}{(t^2-1)^5}\right.\nonumber\\
\left.+\frac{2(c+v)^2(3t^4-14t^2+43)}{(t^2-1)^4}+\frac{4(c+v)^3(t^2-5)}{(t^2-1)^3}
+\frac{2(c+v)^4}{(t^2-1)^2}\right].
\end{eqnarray}

The other autocorrelation functions can be taken from Ref.
\cite{qth04}. For alloy disorder, it is given by \cite{qth04}:
\begin{eqnarray}
\label{Eq:ACFAD}
\langle|U_{\rm AD}(\textit{\textbf{q}})|^2\rangle=x(1-x)u_{\rm al}^2\Omega_0
\frac{B^4b^2}{L}\Bigl[c^4p_0(2b)
+4c^3p_1(2b)+6c^2p_2(2b)+4cp_3(2b)+p_4(2b)\Bigr],
\end{eqnarray}
where $x$ denotes the Ge content of the SiGe alloy, $u_{\rm Al}$ is
the alloy potential. The volume occupied by one alloy atom is given
by $\Omega_0=a_{\rm Al}^3/8$, with $a_{\rm Al}$ the lattice constant
of the alloy. The auxiliary functions $p_l(v)$ ($l=0$ $-$ 4) of the
dimensionless variables $v$ and $b$ used here are defined by
\cite{qth04}
\begin{equation}
\label{Eq:p}
p_l(v)=\frac{b^l}{v^{l+1}}\biggl(1-e^{-v}\sum_{j=0}^l
\frac{v^j}{j!}\biggr).
\end{equation}

The autocorrelation function of interface roughness scattering is
given by \cite{qth04}:
\begin{eqnarray}
\label{Eq:ACFIR} \langle|U_{\rm IFR}(\textit{\textbf{q}})|^2\rangle=\left(\frac{4\pi
e^2}{\epsilon_L}\right)^2B^4\left[n_{\rm D}\left(c^2+2c+2\right)
+\frac{p_{\rm S}B^2}{2}\left(c^4+4c^3+8c^2+8c+4\right)\right]^2
\langle|\Delta_\textit{\textbf{q}}|^2\rangle
\end{eqnarray}
where $n_{\rm D}$ is the depletion charge density.
$\langle|\Delta_\textit{\textbf{q}}|^2\rangle$ is the spectral
distribution of the interface profile usually assumed
\cite{kar90,an82} to be in Gaussian form specified by roughness
amplitude $\Delta$ and correlation length $\Lambda$ as:
\begin{equation}
\label{Eq:IRP} \langle|\Delta_\textit{\textbf{q}}|^2\rangle
=\pi\Delta^2\Lambda^2\exp\left(-\frac{q^2\Lambda^2}{4}\right).
\end{equation}

Deformation potential scattering is a combined effect of lattice
mismatch, which gives rise a strain field $\epsilon$, and interface
roughness \cite{fe95, yin03}. In the previous studies \cite{sa00},
it has always been assumed that the deformation potential
experienced by the holes in the valence band is identical to that
experienced by electrons in the conduction band \cite{fe95} with a
different coupling constant $\Xi$. This assumption is, in fact,
invalid. In particular, while the deformation potential for the
electrons in the conduction band is fixed by a single component
$\epsilon_{zz}$ of the strain field, that for the holes in the
valence band must be fixed by all three diagonal components of
$\epsilon$ \cite{bi74}. Thus, the autocorrelation function for the
deformation potential scattering for the holes needs to be modified
to have the following form \cite{qth04}:
\begin{equation}
\begin{array}{ll}
\label{Eq:ACFDP}
\langle |U_{\rm DP}(\textit{\textbf{q}})|^2\rangle &\displaystyle{=\left(\frac{B^2b^2\alpha
\epsilon_{\parallel}}{2L}\right)^2t^2\Bigl[c^2p_0(b+bt)
+2cp_1(b+bt)
+p_2(b+bt)\Bigr]^2\times}\\
&\displaystyle{
\times\biggl[\frac{3}{2}[b_{\rm
s}(K+1)]^2(1+\sin^4\theta
+\cos^4\theta)+\biggl(\frac{d_{\rm
s}G}{4c_{44}}\biggr)^2\left(1+\frac{\sin^2 2\theta}{4}\right)\biggr]
\langle|\Delta_\textit{\textbf{q}}|^2\rangle,}
\end{array}
\end{equation}
in which $b=kL$, $\epsilon_{\parallel}=(a_{\rm Si}-a_{\rm
Al})/a_{\rm Al}$ is the in-plane strain, $b_{\rm s}$ and $d_{\rm s}$
the shear deformation potential constants. The anisotropy ratio
$\alpha=2c_{44}/(c_{11}-c_{12})$ and the elastic constants
$K=2c_{12}/c_{11}$, $G=2(K+1)(c_{11}-c_{12})$ are deduced from the
elastic stiffness constants of the SiGe alloy $c_{11},\,c_{12}$, and
$c_{44}$.

Similarly, random piezoelectric scattering is a combined effect of
lattice mismatch and interface roughness \cite{qth04}. The other
requirement for this scattering is the piezoelectricity of the
strained SiGe layer, which has recently been found \cite{bra97}.
Because of the interface roughness, the off-diagonal components of
the strain field in the SiGe layer become randomly fluctuating
\cite{qth04, fe95}. Therefore, they induce inside the SiGe layer a
fluctuating density of bulk like piezoelectric charges supplying a
scattering source. The autocorrelation function for piezoelectric
scattering in our HS has been derived by Ref. \cite{qth04} as:
\begin{equation}
\label{Eq:ACFPE} \langle|U_{\rm PE}(\textit{\textbf{q}})|^2\rangle=\left(\frac{3\pi ee_{14}G\alpha
\epsilon_{\parallel}\sin 2\theta}{8\epsilon_{\rm
L}c_{44}}\right)^2 F_{\rm
PE}^2\left(\frac{q}{k}\right)\langle|\Delta_\textit{\textbf{q}}|^2\rangle,
\end{equation}
with $e_{14}$ is the piezoelectric constant of the SiGe alloy
\cite{qth04,bra97}. The piezoelectric form factor $F_{\rm PE}(t)$
appearing in the above equation is given by \cite{qth04}
\begin{eqnarray}
\label{Eq:FFPEW}
F_{\rm PE}(t)=\frac{A^2a}{t+a}\left(1-e^{-2bt}\right)+B^2b
\biggl\{\frac{2c^2t}{t+1}+\frac{4ct}{(t+1)^2}+\frac{4t}{(t+1)^3}+c^2(1-2bt)p_0(b+bt)
\nonumber\\
+2c(1+ct-2bt)p_1(b+bt)
+(1+4ct-2bt)p_2(b+bt)+2tp_3(b+bt)
\nonumber\\
-e^{-2bt}\Bigl[c^2p_0(b-bt)+2cp_1(b-bt)+p_2(b-bt)\Bigr]\biggr\}.
\end{eqnarray}

The background impurity scattering examined in some previous studies
\cite{kar90, kar91} is not considered here. The reason is that our
HS has only an impurity sheet separated from the well by a spacer of
thickness $L_{\rm S}=120$\AA, but no intentional background impurity
\cite{pos04, col97}. Thus, the Eqs. (\ref{Eq:ACFMI}),
(\ref{Eq:ACFAD}), (\ref{Eq:ACFIR}), (\ref{Eq:ACFDP}), and
(\ref{Eq:ACFPE}) supply all needed for specifying the relaxation
time using the Eq. (\ref{Eq:tau}).

\section{Numerical results and discussions}

\subsection{Comparison to experiment}
We now compare calculated $S^{\rm d}$ to some experimental data for
the thermopower $S$ ($T<0.5{\rm K}$) of the 2D hole gas in a
$p-$type Si/Si$_{0.88}$Ge$_{0.12}$ HS reported on the Fig. 6 of the
Ref. \cite{pos04}. This HS, which is the sample CVD191 used in the
Refs. \cite{pos04,col97}, is composed by a strained
Si$_{0.88}$Ge$_{0.12}$ layer of thickness $L$ grown on a Si
substrate. A spacer of thickness $L_{\rm S}$ followed by an impurity
sheet of thickness $L_{\rm I}$ is placed on the top of the well. At
$p_{\rm S}=2.7\times 10^{11}\rm{cm}^{-2}$, the hole mobility $\mu$
is $15000\rm{cm}^2/\rm{Vs}$, while the thermopower $S$ is given
\cite{pos04} by the solid squares in the Fig. 1. As $T\leq 0.5{\rm
K}$, $S$ is approximately linear, reflecting the domination of the
diffusion thermopower $S^{\rm d}$ calculated by Eq. (\ref{Eq:p1})
using a \textit{phenomenological} expression for $p$ with five
fitting parameters but no calculations starting from the microscopic
level \cite{pos04}.

\begin{figure}[ht]
  \begin{center}
  \vspace{-6mm}
  \includegraphics[width=9.5cm]{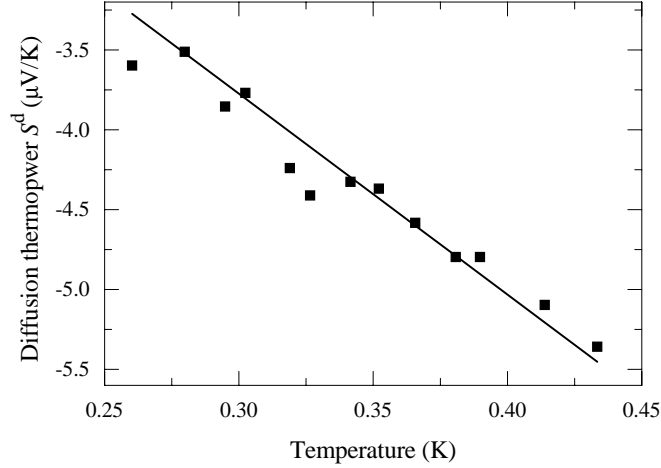}
  \vspace{-4mm}
  \caption{Measured thermopower $S$ (squares) reported in Ref. \cite{pos04}
      and calculated diffusion
  thermopower $S^{\rm d}$ (solid line) of the 2D hole gas vs
  temperature $T$.  Calculation parameters are
  $\Delta=1.3$~\AA~, $\Lambda=97$~\AA, and $n_{\rm I}=3.0\times 10^{18}{\rm cm}^{-3}$.}
  \end{center}
\end{figure}

Our calculations, on the other hand, are based on experimental
parameters using $\Delta$ and $\Lambda$ as fitting parameters. These
parameters are: $L=400$\AA, $L_{\rm S}=120$\AA, $L_{\rm I}=300$\AA,
$x=0.12$, $n_{\rm I}=3.0\times 10^{18}{\rm cm}^{-3}$, $m^*=0.29m_e$
\cite{pos04, col97}. The finite barrier is chosen to be
$V_0(x)=0.74x=0.089$ eV as in the Ref. \cite{peo86} while the alloy
disorder potential $u_{\rm Al}=0.30~{\rm eV}$ as in the Ref.
\cite{lai93}. Following the Refs. \cite{qth04,lai93}, the other
parameters are: $a_{\rm Si}=5.43$~\AA, $a_{\rm Ge}=5.658$~\AA~,
$a_{\rm Al}=5.455$ \AA, $n_{\rm D}=5.0\times 10^{10}~{\rm cm}^{-2}$,
$\epsilon_{\rm L}=12.192$, $b_{\rm s}=-2.804~{\rm eV}$, $d_{\rm
s}=-5.240~{\rm eV}$, $c_{11}=16.15\times 10^{10}~{\rm Pa}$,
$c_{12}=6.203\times 10^{10}~{\rm Pa}$, $c_{44}=7.821\times
10^{10}~{\rm Pa}$ and $e_{14}=0.956\times 10^{-2}~{\rm C/m^2}$.
Detail discussions in choosing parameters are available in Ref.
\cite{qth04}. As a result, calculations with $\Delta=1.3$~\AA~ and
$\Lambda=97$~\AA~give $\mu=14709~\rm{cm}^2/\rm{Vs}$ and $S^{\rm
d}/T$ is $-12.60~ \mu \rm{V}/\rm{K}^2$, providing the best fit to
the reported data, as seen on the Fig. 1.

It has been suggested both experimentally \cite{fe95, yin03, xie94}
and theoretically \cite{quang05,yam96} that $\Lambda\sim100$~\AA~
while $\Delta$ varies from 1~\AA~ to 20~\AA. In fact, the roughness
amplitude $\Delta=1.3$ \AA~is small comparing to those normally used
\cite{kar90, kar91, qth04, sa00, kub05}, thus it is necessary to
propose an interpretation for this value. It has been pointed out
\cite{fe95,yin03} that a strained layer, as thinner than a critical
thickness \cite{peo85}, prefer to relax by buckling which gives
raise interface roughness. Consequently, $\Delta$ is found to depend
strongly on the strained SiGe layer thickness \cite{yin03}, the Ge
content of the SiGe layer \cite{yin03, xie94}, and the cap layer
thickness \cite{yin03}. In our HS \cite{pos04,col97}, the SiGe layer
with $L=400$ \AA~ is thick, the Ge content $x=0.12$ is small, and
the cap layer is thick ($L_{\rm S}+L_{\rm I}= 420$~\AA). All of
them, interestingly, supply a small $\Delta$ \cite{yin03, xie94}.
Further, there exist in the literature many studies supporting
$\Delta$ at the same order. Very small roughness amplitudes
($\Delta\simeq 1$\AA) of Si/SiGe interfaces have been seen
experimentally in the Ref. \cite{xie94}. The same $\Delta$ were also
used theoretically, including $\Delta = 2$~\AA~ in the Ref.
\cite{quang05}, and $\Delta = 1.78$~\AA~ in the Ref. \cite{yam96}.

\subsection{Scattering mechanisms and the diffusion thermopower}

In order to examine the strengths of the existing scatterings in our
HS, the hole mobilities limited by separated and combined
scatterings are given on the Fig. 2 using the expression
$\mu=e\tau(E_{\rm F})/m^*$ within the linear transport theory
\cite{an82}. The parameters of the Fig. 1 are used here. The Fig. 2
indicates that in our HS, piezoelectric scattering is weak and alloy
disorder is of minor important on the whole range of $p_{\rm S}$. As
$p_{\rm S}\geq 1.5\times 10^{11}{\rm cm}^{-2}$, deformation
potential and interface roughness are the dominant scatterings while
in the region $p_{\rm S}\leq 1.5\times 10^{11}{\rm cm}^{-2}$, remote
impurity dominates. There are two reasons for the major importance
of interface roughness, in spite of small $\Delta$. First, since the
barrier is small, the hole wave function penetrates deeply into the
substrate. Consequently, the hole density at the barrier is finite
instead of zero for the infinite barrier, thus strengthening
interface roughness scattering \cite{qth04}. Next, the finite
barrier suppresses both alloy disorder and deformation potential
scatterings, as showed by the Ref. \cite{qth04}.

\begin{figure}[ht]
  \begin{center}
  \vspace{-6mm}
  \includegraphics[width=9.5cm]{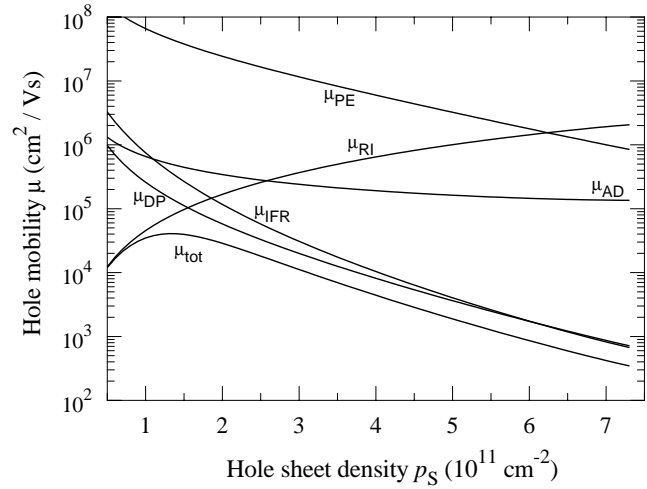}
    \vspace{-5mm}
  \caption{Partial and total mobilities of the 2D hole gas
  in Si/Si$_{0.88}$Ge$_{0.12}$ HS vs hole density $p_{\rm S}$.
  $\mu_{\rm RI}$, $\mu_{\rm AD}$,
$\mu_{\rm IFR}$, $\mu_{\rm DP}$, $\mu_{\rm PE}$ are the partial
mobilities limited by  remote impurity, alloy disorder, interface
roughness, deformation potential, and piezoelectric, respectively.
The total mobility $\mu_{\rm tot}$ is limited by all of the
scatterings.}
  \end{center}
\end{figure}

Next, we examine the partial and total $S^{\rm d}/T$ which are
plotted vs. $p_{\rm S}$ on the Fig. 3 with the same parameters of
Fig. 2. The notations $S^{\rm d}_{\rm RI}$, $S^{\rm d}_{\rm AD}$,
$S^{\rm d}_{\rm IFR}$, $S^{\rm d}_{\rm DP}$, $S^{\rm d}_{\rm PE}$
represent the partial diffusion thermopowers due to remote impurity,
alloy disorder, interface roughness, deformation potential, and
piezoelectric scatterings, respectively. As mentioned above, the
total $S^{\rm d}$ is a combination of the partial components,
weighted by the corresponding scattering strengths. An examination
of Fig. 3 reveals that $S^{\rm d}$ changes its sign at $p_{\rm
S}\simeq 1.8\times 10^{11}{\rm cm}^{-2}$, which is smaller than that
reported for 2D electron gas \cite{kar90, kar91}. On the whole range
of $p_{\rm S}$, $S^{\rm d}_{\rm PE}$ and $S^{\rm d}_{\rm IFR}$ are
almost independent of $p_{\rm S}$. Differently, $S^{\rm d}_{\rm DP}$
is very negative and depends strongly on $p_{\rm S}$ as $p_{\rm
S}\leq 2.0\times 10^{11}{\rm cm}^{-2}$ before becoming independent
of $p_{\rm S}$ as $p_{\rm S}\geq 4.0\times 10^{11}{\rm cm}^{-2}$.
While $S^{\rm d}_{\rm AD}$ is very small and may be neglected,
$S^{\rm d}_{\rm RI}$ is the only one having large positive values.

\begin{figure}[ht]
  \begin{center}
  \vspace{-6mm}
  \includegraphics[width=9.5cm]{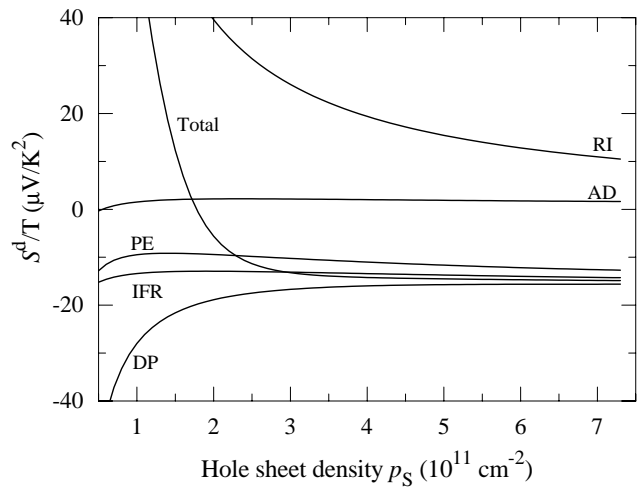}
    \vspace{-5mm}
  \caption{Partial and total $S^{\rm d}/T$
of the Si/Si$_{0.88}$Ge$_{0.12}$ HS in Fig. 2 as functions of
$p_{\rm S}$ with the parameters taken from Fig. 2. RI, AD, IFR, DP,
PE are the abbreviations of remote impurity, alloy disorder,
interface roughness, deformation potential, and  piezoelectric
scatterings, respectively.}
  \end{center}
\end{figure}

Now we turn to another interesting issue: the possibility of
changing in sign of $S^{\rm d}$ when the SiGe layer thickness $L$
changes. While the change in sign of $S^{\rm d}$ in $n-$ type
Si-MOSFET's as the carrier density varies has been addressed
\cite{kar90, kar91}, no discussion on the dependence of $S^{\rm d}$
on $L$ has been given. A recent study \cite{kub05}, on the other
hand, suggests that the $S^{\rm d}$ of an $n-$type GaAs quantum well
can change its sign as the well thickness $L$ changes. This
possibility is a consequence of a strong piezoelectric scattering in
the quantum well made by GaAs material with a large piezoelectric
constant $e_{14}$. It may be interesting to figure out if there is
such a possibility in a SiGe alloy with a smaller $e_{14}$?

\begin{figure}[ht]
  \begin{center}
  \vspace{-4mm}
  \includegraphics[width=9.5cm]{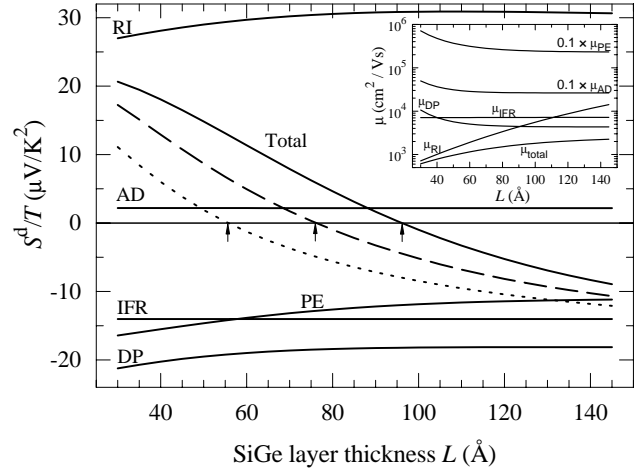}
  \vspace{-4mm}
  \caption{Plot of partial and total $S^{\rm d}/T$ of a
Si/Si$_{0.88}$Ge$_{0.12}$ HS as functions of the SiGe layer
thickness $L$ with $L_{\rm S}=100$\AA~ (solid line). Dashed and
dotted lines give the total $S^{\rm d}/T$ for $L_{\rm S}=125$\AA~
and $L_{\rm S}=150$\AA, respectively. The hole mobilities limited by
separated and combined scatterings of this HS are given on the
inset.}
  \end{center}
\end{figure}

To answer this question, we look for experimentally attainable
parameters which allow our HS to exhibit a change in sign of $S^{\rm
d}$ as $L$ changes. The following parameters are kept: $p_{\rm
S}=2.7\times 10^{11} {\rm cm}^{-2}$, $L_{\rm I}=300$~\AA, $x=0.12$.
The others are chosen for a weaker remote impurity scattering:
$n_{\rm I}=1\times 10^{18} {\rm cm}^{-3}$, $L_{\rm S}=100$~\AA,
$\Delta=3$ \AA, $\Lambda=100$ \AA. The partial and total $S^{\rm
d}/T$ are plotted vs. $L$ on the Fig. 4, which shows that the total
diffusion thermopower changes its sign at the SiGe layer critical
thickness $L_{\rm C}\simeq 96$~\AA.

For more information, the inset of the Fig. 4 provides the hole
mobilities of the HS limited by separated and combined scatterings.
It can be seen that while alloy disorder and piezoelectric
scatterings are small, the others are comparable. Thus, remote
impurity, deformation potential and interface roughness are
important scatterings determining the total diffusion thermopower.
Since nothing but $S^{\rm d}_{\rm RI}\gg 1$ (Fig. 3), we can adjust
$L_{\rm C}$ by changing the strength of remote impurity scattering.
There are several ways to do that, including changing the spacer
thickness or impurity density. Indeed, our calculations reveal that
$L_{\rm C}$ depends strongly on $L_{\rm S}$. For illustrations, the
dashed line on the Fig. 4 shows that for $L_{\rm S} =125$~\AA,
$L_{\rm C}\simeq 76$~\AA. If the spacer is wider, ($L_{\rm
S}=150$~\AA), the critical thickness is even much lower: $L_{\rm
C}\simeq 55$~\AA~ (the dotted line).

\section{Conclusion}
In conclusion, we present a theoretical study of the diffusion
thermopower $S^{\rm d}$ in a $p-$type Si/Si$_{1-x}$Ge$_x$ lattice
mismatched HS at low temperature and zero magnetic field. In the HS,
deformation potential, alloy disorder, and piezoelectric scatterings
are examined in comparing to the conventional scatterings. The
calculated diffusion thermopower is in good agreement with a recent
experiment. Further, $S^{\rm d}$ is found to depend strongly on the
SiGe layer thickness $L$, and changes its sign as $L$ across a
critical thickness $L_{\rm C}$. The possible parameters which can
affect $L_{\rm C}$ is also proposed. Deformation potential is a
dominant scattering, making an important contribution to $S^{\rm
d}$. On the other hand, piezoelectric is weak while alloy disorder
has a very small partial $S^{\rm d}$. Changing the hole density, we
find a sign change of $S^{\rm d}$ at a smaller hole density
comparing to that reported previously for 2D electron gases.

\section*{Acknowledgments}
The authors would like to thank Professor J. C. Maan, University of
Nijmegen, The Netherlands for supplying the data used in the Fig. 1
of this paper. They also thank the referees for requiring them to be
more precise in preparing this manuscript.


\begin{thebibliography}{00}
\bibitem{pos04} C. Possanzini, R. Fletcher, M. Tsaousidou, P. T. Coleridge,
R. L. Williams, Y. Feng, and J. C. Maan, Phys. Rev. B {\bf 69},
195306 (2004).
\bibitem{kar90} V. K. Karavolas, M. J. Smith, T. M. Fromhold,
P. N. Butcher, B. G. Mulimani, B. L. Gallagher and J. P. Oxley,
J. Phys.: Condens. Matter {\bf 2}, 10401 (1990).
\bibitem{kar91} V. C. Karavolas, and P. N. Butcher, J. Phys.:
Condens. Matter {\bf 3}, 2597 (1991).
\bibitem{qth04} D. N. Quang, V. N. Tuoc, T. D. Huan, and P. N. Phong, Phys.
Rev. B {\bf 70}, 195336 (2004).
\bibitem{sa00} M.A. Sadeghzadeh, A.I. Horrell, O.A. Mironov, E.H.C.
Parker, T.E. Whall, and M.J. Kearney, Appl. Phys. Lett. {\bf 76},
2568 (2000), and references therein.
\bibitem{kub05} S. S. Kubakaddi and K. R. Usharani, Physica E {\bf
25}, 497 (2005).
\bibitem{fle97} R. Fletcher, V. M. Pudalov, Y. Feng, M. Tsaousidou and P. N. Butcher,
Phys. Rev. B {\bf 56}, 12422 (1997).
\bibitem{fle02} R. Fletcher, M. Tsaousidou, P. T. Coleridge,
Y. Feng and Z. R. Wasilewski, Physica E {\bf 12}, 478 (2002).
\bibitem{an82} T. Ando, A. B. Fowler, and F. Stern, Rev. Mod. Phys. {\bf 54}, 437 (1982).
\bibitem{oku89} Y. Okuyama and N. Tokuda, Phys. Rev. B {\bf 40}, 9744 (1989).
\bibitem{col97} P. T. Coleridge, R. L. Williams, Y. Feng, and P.
Zawadzki, Phys. Rev. B {\bf 56}, 12764 (1997).
\bibitem{an82b} T. Ando, J. Phys. Soc. Jpn. {\bf 51}, 3893 (1982);
{\it ibid.} {\bf 51}, 3900 (1982).
\bibitem{quang05} D. N. Quang, V. N. Tuoc, N. H. Tung, N. V. Minh,
and P. N. Phong, Phys. Rev. B {\bf 72},
245303 (2005).
\bibitem{sch90} E. F. Schubert, J. M. Kuo, R. F. Kopf,
H. S. Luftman, L. C. Hopkins, and N. J. Sauer, J. Appl. Phys. {\bf
67}, 1969 (1990).
\bibitem{efr90} A. L. Efros, F. G. Pikus, and G. G. Samsonidze, Phys. Rev. B {\bf 41},
8295 (1990).
\bibitem{quang99} D. N. Quang and N. H. Tung, phys. stat. sol. (b) {\bf 207},
111 (1998).
\bibitem{fe95} R. M. Feenstra and M. A. Lutz, J. Appl. Phys. {\bf 78}, 6091 (1995).
\bibitem{yin03} H. Yin, R. Huang, K. D. Hobart, J. Liang, Z. Suo,
S. R. Shieh, T. S. Duffy, F. J. Kub, J. C. Sturm, J. Appl. Phys.
{\bf 94}, 6875 (2003).
\bibitem{bi74} G.L. Bir and G.E. Pikus, \emph{Symmetry and Strain Induced Effects in
Semiconductors} (Wiley, New York, 1974).
\bibitem{bra97} G. Braithwaite, N. L. Mattey, E. H. C. Parker,
T. E. Whall, G. Brunthaler, and G. Bauer, J. Appl. Phys.
{\bf 81}, 6853 (1997).
\bibitem{peo86} R. People and J. C. Bean, Appl. Phys. Lett. {\bf 48}, 538 (1986).
\bibitem{lai93} B. Laikhtman and R.A. Kiehl, Phys. Rev. B \textbf{47}, 10 515 (1993).
\bibitem{xie94} S. M.
Goodnick, D. K. Ferry, C. W. Wilmsen, Z. Liliental, D. Fathy, and O.
L. Krivanek, Phys. Rev. B {\bf 32}, 8171 (1985).
\bibitem{yam96} S. Yamakawa, H. Ueno, K. Taniguchi, C. Hamaguchi,
K. Miyatsuji, K. Masaki, and U. Ravaioli, J. Appl. Phys. {\bf 79},
911 (1996).
\bibitem{peo85} R.
People and J. C. Bean, Appl. Phys. Lett. {\bf 47}, 322 (1985); {\it
ibid.} {\bf 49}, 229 (1986).

\end{thebibliography}
\end{document}